\documentstyle[aps,prl,twocolumn,epsf]{revtex}

\newcommand{\Artanh}{{\rm Artanh}\,}

\newcommand{\B}{{\rm B}}
\newcommand{\I}{{\rm I}}
\newcommand{\T}{{\rm T}}

\renewcommand{\S}{{\rm S}}

\begin{document}

\draft

\title{Kinetic Freeze-out and Radial Flow in 11.6 $A$~GeV Au+Au Collisions}
\author{Harald Dobler$^a$ \and Josef Sollfrank$^a$ \and 
        Ulrich Heinz$^{b,}$\thanks{On leave from Institut f\"ur
        Theoreti\-sche Physik, Universit\"at Regensburg; email address: 
        Ulrich.Heinz@cern.ch}}
\address{$^a$Institut f\"ur Theoretische Physik, Universit\"at
  Regensburg, D-93040 Regensburg, Germany\\
  $^b$Theoretical Physics Division, CERN, CH-1211 Geneva 23, Switzerland}
\date{\today}
\maketitle

\begin{abstract}
We study the kinetic freeze-out conditions of hadrons in Au+Au collisions
at 11.6 $A$\,GeV/$c$ using different pa\-ra\-me\-tri\-za\-tions of an 
expanding thermal fireball. We take into account the available double 
differential momentum spectra of a variety of particle species, covering 
a large fraction of the total momentum space. The overall fit to the 
data is very good and indicates a relatively low kinetic freeze-out 
temperature of about $90$ MeV with an average transverse expansion 
velocity at midrapidity of about $0.5\,c$.
\end{abstract}

\pacs{PACS numbers: 25.75.-q \and 25.75.Ld \and 24.10.Pa}

When studying the nuclear phase diagram in extreme regions of 
temperature and density with heavy-ion collisions one is faced with 
the difficulty that, due to the li\-mi\-ted size and strong dynamics of 
the fireballs created in such collisions, full (global) thermodynamic 
equilibrium and the thermodynamic (infinite volume) limit can never 
be reached. The best one can hope for is a state of {\em local} 
thermal equilibrium in a hydrodynamically expanding environment, 
with thermodynamic homogeneity volumes of several 100 to 1000 fm$^3$ 
\cite{Heinz99}. If sufficient local equilibration is achieved
one can try to extrapolate from the experimental findings to 
the thermodynamic limit. The most exciting prospect of such an 
endeavour is the experimental confirmation of a deconfinement phase 
transition in hot and dense nuclear matter \cite{Harris96}.

In order to reach this goal an extensive experimental program has 
been launched studying heavy-ion collisions at various beam energies. 
The onset of a new phase of nuclear matter would probably be most 
clearly seen in a study of different observables as functions of 
beam energy. For this purpose it is necessary to carefully analyze
the data at all available beam energies. Here we present a study of 
the experimental particle spectra from Au+Au collision at a beam 
momentum of 11.6\,$A$\,GeV/$c$ measured by several groups 
\cite{E866,E877,E891} at the Brookhaven AGS. The goal is to find 
a simple, but realistic parametrization of the freeze-out stage in 
these collisions. The extracted freeze-out parameters, especially 
the temperature and the mean transverse flow velocity, will be 
compared with other collision systems and with collisions at different
energies. This should help to understand the gross features of the 
collision dynamics at ultrarelativistic energies.

For the description of particle production we start from the 
formalism of Cooper and Frye \cite{CooperFrye} which describes 
the single-particle spectrum as an integral over a freeze-out
hypersurface, thus summing the contributions from all space-time 
points at which the particles decouple from the fireball:
 \begin{equation}
 \label{1}
   E\frac{d^3N}{d^3p} = \frac{g}{(2\pi)^3} 
   \int_{\Sigma_{\rm f}} p^\mu d\sigma_\mu(x)\, f(x,p) \:.
 \end{equation}
Here $g$ is the degeneracy factor and $f(x,p)$ the momentum 
distribution at space-time point $x$. In thermal models one
uses for $f(x,p)$ a thermal equilibrium distribution and determines 
$\Sigma_{\rm f}$ by a freeze-out criterium for thermal decoupling 
\cite{Schnedermann}. For the low temperatures discussed below the 
Boltzmann approximation is sufficient, but we allow for a
space-time dependence of the temperature $T$, the chemical 
potential $\mu$, and the flow velocity $u^\mu$:
 \begin{equation} 
 \label{dist}
   f(x,p) = \exp\left(-\frac{p \cdot u(x) - \mu(x)}{T(x)}\right) \: .
 \end{equation}

Our choice of the geometry of the freeze-out surface $\Sigma_{\rm f}$ 
and of the space-time dependence of the parameters reflects a 
compromise between relative simplicity of the model and importance 
of its dynamical ingredients. Focussing on central collisions, we 
assume azimuthal symmetry of the spatial geometry and momentum 
distributions. We further assume boost-invariant collective dynamics 
along the longitudinal ($z$) direction \cite{Bjorken}. This was shown 
before \cite{Stachel96} to give a better representation of the 
measured rapidity spectra than a static ``Hagedorn'' fireball. Also, 
a static fireball yields a much stronger rapidity dependence of the 
transverse slope parameters, $T_{\rm slope} = T/\cosh(y-y_{\rm mid})$, 
than observed \cite{E866,E877,E891}. The assumption of longitudinal 
Bjorken flow suggests to use longitudinal proper time 
$\tau = \sqrt{t^2-z^2}$ and space-time rapidity $\eta = 
\Artanh (z/t)$ as suitable variables in the $t$-$z$-plane; the 
transverse radial coordinate is denoted by $r_\perp$. Midrapidity 
is defined to be at $\eta = 0$. Longitudinal boost-invariance
of the flow field is implemented by the choice $u^\mu(x) = 
\gamma(1,v_\perp(x)\bbox{e}_r,v_z(x))$ with $v_z(\tau,r_\perp,\eta) = 
\tanh \eta$. In transverse direction we take a linear flow rapidity
profile, $\tanh v_\perp = \rho(\eta){r_\perp\over R_0}$, where $R_0$ 
is the transverse radius at midrapidity and $\rho(\eta)$ will be 
specified below.

The geometry of the freeze-out hypersurface $\Sigma_{\rm f}$ is 
fixed as follows: In the time direction we take a surface
of constant proper time, $\tau=\tau_0$. In $\eta$-direction 
the freeze-out volume extends only to a maximum space-time 
rapidity $\eta_{\max}$; this breaks longitudinal boost-invariance,
as required by the finite available total energy. In the transverse 
direction the boundary is given by $R(\eta)$ which may depend on 
$\eta$. We consider two choices:
 \begin{eqnarray}
   R(\eta) &=& R_0\cdot\Theta\left(\eta_{\max}^2 -\eta^2\right) \,,
 \label{cylinder}\\
   R(\eta) &=& R_0\cdot
   \sqrt{1 - {\eta^2\over\eta_{\max}^2}}\,.
 \label{ellipsoid}
 \end{eqnarray}
The first choice (\ref{cylinder}) describes a cylindrical fireball
in the $\eta$-$\bbox{r}_\perp$-space. For 200 $A$ GeV S+S collisions 
at the SPS such a picture was shown to work well \cite{Schnedermann}, 
but for the lower AGS beam energy this can no longer be expected. 
An elliptic or cigar shape as given in Eq.~(\ref{ellipsoid}) should 
be more appropriate \cite{Nix}.

Having specified the freeze-out geometry and the distribution function 
(\ref{dist}) at freeze-out we obtain the following thermal single-particle 
spectrum:
 \begin{eqnarray}
   \frac{dN}{m_\T dm_\T  dy}
   &=&
   \frac{g}{2\pi}\,m_\T\,\tau_0
   \int_{-\eta_{\max}}^{+\eta_{\max}} d\eta\,\cosh(y-\eta)
 \nonumber\\
   &\times&
   \int_0^{R(\eta)} r_\perp\,dr_\perp \, 
   \I_0\left(\frac{p_\T\sinh\rho(r_\perp)}{T(x)}\right)
 \label{therm}\\
   &\times&
   \exp\left(\frac{\mu(x)-m_\T\cosh(y{-}\eta)\cosh\rho(r_\perp)}
                  {T(x)}\right) .
 \nonumber 
 \end{eqnarray}
Before comparing with data we must still add the contributions 
from resonance decays. This is done following 
Refs.~\cite{Sollfrank,RajuVeng}, again using thermal distributions
(\ref{therm}) for the resonances. This procedure implies the 
assumption of full chemical equilibrium, including strange particles. 
As we will see this assumption is not very well justified, but 
sufficient for estimating resonance feeddown contributions. We only 
include resonances with masses up to the $\Delta(1232)$ resonance. 
Tests showed that at the low freeze-out temperatures discussed below 
the inclusion of higher resonances has very little influence 
on the results.

At this point only $T(x)$ and $\mu(x)$ remain to be spe\-ci\-fied. The 
chemical potential of a hadron with baryon number $B$ and strangeness 
$S$ is fixed by the baryonic and strange chemical potentials as  
$\mu(x) = B\cdot\mu_\B(x) + S \cdot\mu_{\rm S}(x)$ (we neglect small 
isospin asymmetry effects \cite{isospin}). We have investigated two 
scenarios. In the first we set for simplicity $T$, $\mu_\B$, and 
$\mu_{\rm S}$ constant along the freeze-out hypersurface. In the 
second we select a Gaussian $\eta$-dependence of the freeze-out
temperature,
 \begin{equation}
 \label{eq:Gauss}
   T(\eta) = T_0\, e^{-\left(\frac{\eta}{\Delta\eta}\right)^2}\, ,
 \end{equation}
and fix $\mu_\B(\eta)$ by requiring freeze-out at constant baryon 
density. This is reasonable because in the range of $T$ and $\mu_\B$ 
discussed below meson-meson reactions occur much less frequently 
than meson-baryon or baryon-baryon reactions, making the baryon 
density the relevant control parameter for freeze-out. The 
strangeness chemical potential $\mu_{\rm S}$ is obtained as a 
function of $T(\eta)$ and $\mu_\B(\eta)$ from the condition of 
local strangeness neutrality \cite{isospin}. The 
thermal fit parameters in the model are thus $\mu_\B(\eta{=}0)$ and $T_0$, 
the baryon chemical potential and temperature at midrapidity, and the 
width $\Delta\eta$ of the temperature profile (\ref{eq:Gauss}).

%
 \begin{figure}
  \epsfxsize 65mm \epsfbox{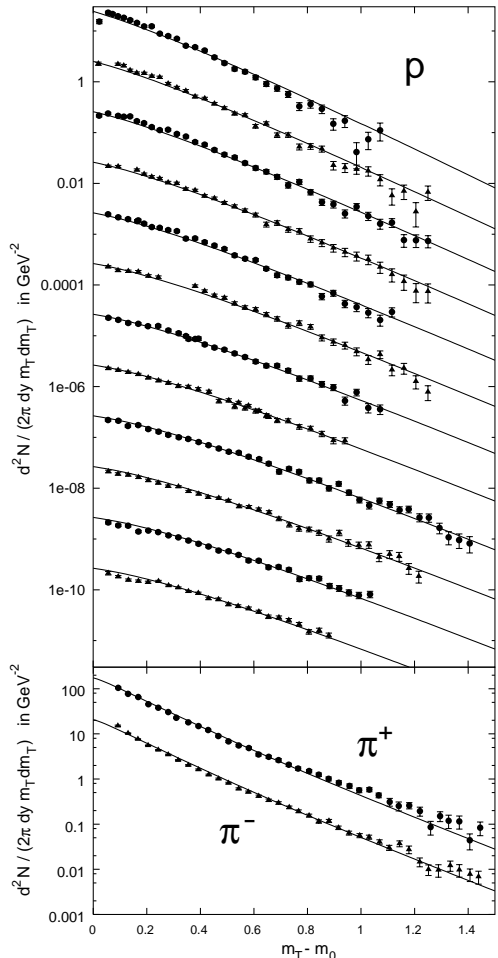}
  \caption{
      Proton and pion $m_\T$-spectra from E866 \protect\cite{E866}.
      The proton spectra are given in rapidity bins of width $0.1$;
      the top spectra correspond to $0.5{\leq}y_{\rm lab}{\leq}0.6$,
      the bottom ones to $1.6{\leq}y_{\rm lab}{\leq}1.7$. (Midrapidity
      is at $y_{\rm lab}$=1.6.) Successive spectra are scaled down by 
      a factor 10. The pion spectra are given for 
      $1.4{\leq}y_{\rm lab}{\leq}1.6$; the $\pi^-$ spectrum is 
      divided by $10$. Solid lines: best fit with Model 3.
      \label{F1}}
 \end{figure}
%

\begin{figure}
  \epsfxsize 80mm \epsfbox{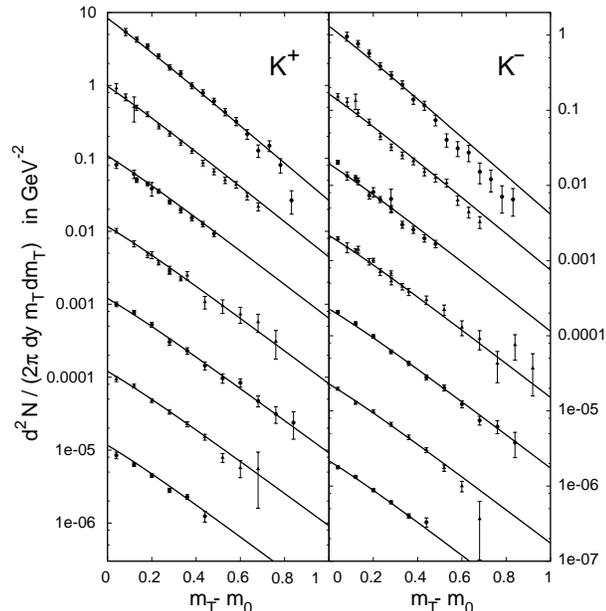}
  \caption{
      Kaon $m_\T$-spectra from E866 \protect\cite{E866}, 
      in rapidity bins of width $0.2$. Successive spectra are 
      scaled down by a factor 10. The top spectra 
      correspond to $0.6{\leq}y_{\rm lab}{\leq}0.8$, the bottom 
      ones to $1.8{\leq}y_{\rm lab}{\leq}2.0$. Solid lines: best 
      fit with Model 3.
 \label{F2}}
 \end{figure}

Boost-invariant longitudinal flow combined with a cylindrical fireball
geometry leads to nearly rapidity-independent transverse mass spectra
near midrapidity. (For the present data midrapidity corresponds to 
$y_{\rm lab}{=}1.6$.) This contradicts the measurements which show a 
clear rapidity dependence of the average transverse momentum 
$\langle p_T \rangle$ and of the transverse slope parameters also
near midrapidity, especially for the heavier protons and $\Lambda$'s
\cite{E866,E877,E891} (see also  Figs.~\ref{F1},\ref{F2} and the 
discussion below). To describe the data one must therefore either 
reduce the transverse flow or the temperature as one moves away from 
midrapidity. Having investigated a wide range of possibilities, we 
here report on the results obtained by combining in various ways the 
following relatively simple options: Cylindrical (\ref{cylinder}) vs. 
elliptic (\ref{ellipsoid}) transverse geometry; constant vs. 
$\eta$-dependent (\ref{eq:Gauss}) freeze-out temperature; and 
a constant ($\rho(\eta){\equiv}\rho_0$) vs. an $\eta$-dependent 
($\rho(\eta){=}\rho_0\sqrt{1{-}(\eta^2/\eta_{\rm max}^2)}$)
transverse flow gradient. The corresponding model parameters 
are listed in Table~\ref{T1}. 

%
\begin{table}
\begin{tabular}{cl}
symbol & description \\
\hline
$V\sim\tau_0 R_0^2$ & fireball volume \\
$\eta_{\max}$ & maximal longitudinal flow rapidity\\
$\rho_0$ & maximal transverse flow rapidity\\
$\mu_{\rm B}(\eta=0)$ & baryon chemical potential at midrapidity\\
$T_0$ & temperature at midrapidity\\
$\Delta\eta$ & width of the longitudinal temperature profile\\
\end{tabular}
\caption{Model parameters. Note that $\tau_0$ and $R_0$ can
  be scaled out from the integrals in (\ref{therm}), leaving only the 
  combination $\tau_0 R^2_0$ as an independent fit parameter. It is 
  proportional to the Lorentz invariant proper fireball volume 
  $V = \int_{\Sigma_{\rm f}} u^\mu d\sigma_\mu$ which is quoted in 
  Table~\protect\ref{T2}.
  \label{T1}}
\end{table}
%

{\em Model 1} combines cylindrical geometry with constant temperature
and transverse flow; as already mentioned it does not produce a very 
good fit. The parameters corresponding to the smallest $\chi^2$/DOF 
are listed in the first column of Table~\ref{T2}. The quoted value of 
$\chi^2$/DOF=6.64 was obtained after allowing for different overall 
normalizations for the various particle species than predicted by 
chemical equilibrium \cite{fn1}. The corresponding normalization 
factors ${\cal N}_i$ are also given; their deviation from 1 indicates 
the departure from chemical equilibrium at the point of thermal 
freeze-out. Effects of up to 25{\%} are seen and should be expected 
in view of the finding in \cite{Rafelski,Braun,Stachel96} that chemical 
freeze-out at the AGS occurs already at $T_{\rm chem} \simeq 130$ MeV.

Model 1 fails in two characteristic ways: (1) The rapidity 
spectra for heavy particles are too flat near midrapidity and too 
steep near $y=\pm \eta_{\max}$. (2) The slopes of the $m_\T$-spectra are 
essentially constant in the region $\vert y\vert \leq \eta_{\max}$
while the data show a significant steepening of the spectra
already closer to midrapidity. Both failures are due to the 
boost-invariance of Model 1 in the region 
$\vert\eta\vert \leq \eta_{\max}$.

Due to the well-known weak sensitivity of the rapidity spectra to 
changes in the temperature \cite{Schnedermann} it is hard to 
improve on them by simply allowing the freeze-out temperature to 
depend on $\eta$; with the parametrization (\ref{eq:Gauss}) we found 
it to be impossible. A smooth reduction of the effective tranverse 
flow as described above works much better, in particular if the
simultaneous improvement on 
\newpage

\noindent
the rapidity dependence of the transverse slopes is taken into 
account. Nevertheless, the proton and $\Lambda$ rapidity distributions 
are still a bit too flat in the middle and too steep at the edges. 

%
\begin{table}
\begin{tabular}{cccc}
parameter & Model 1 & Model 2 & Model 3 \\
\hline
$V / (10^4\,{\rm fm}^3)$  &  1.74  &  1.74  &  1.82\\
$\eta_{\max}$  &    1.05  &  1.66  &  1.71\\
$\rho_0$  &  0.726  &  0.793  &  0.792\\
$\mu_\B(\eta=0) / {\rm MeV}$  &  548  &  549  &  536\\
$T_0 / {\rm MeV}$  &  90.5  & 91.6  &  93.3\\
$\Delta\eta$  &  $\infty$  &  $\infty$  &  3.58\\
\hline
$\mu_\S(\eta=0) / {\rm MeV}$  &  61.1  &  62.7  &  59.5\\
$\langle v_\perp \rangle(\eta=0)$  &  0.449  &  0.484  &  0.483\\
\hline
${\mathcal N}_{\pi^+}$          &  0.924  &  0.920  &  0.920 \\
${\mathcal N}_{\pi^-}$          &  1.101  &  1.090  &  1.078 \\
${\mathcal N}_{{\rm K}^+}$  &  1.218  &  1.239  &  1.268 \\
${\mathcal N}_{{\rm K}^-}$  &  0.887  &  0.915  &  0.914 \\
${\mathcal N}_{{\rm p}}$      &  0.972  &  0.978  &  0.972 \\
${\mathcal N}_\Lambda$      &  1.115  &  1.078  &  1.206 \\
\hline
$\chi^2/$DOF  &  6.64  &  3.16  &  2.40\\
\end{tabular}
\caption{Fit results for the different models (see text for their 
  description).
  \label{T2}}
\end{table}
%

Surprisingly, a very good fit of both the rapidity and all the 
$m_\T$-spectra can be obtained by simply reducing the transverse
size of the fireball according to (\ref{ellipsoid}), without changing
the freeze-out temperature or transverse flow gradient ({\em Model 2}).
The reduced transverse radius in backward and forward rapidity regions
not only leads to a faster and smoother decrease of the proton and
$\Lambda$ rapidity distributions, but the implied reduced {\em average} 
transverse flow away from midrapidity allows for an excellent fit
to the steepening $m_\T$-spectra. Combining Eq. (\ref{ellipsoid}) with 
an $\eta$-dependence of the transverse flow gradient overdoes it: the 
rapidity spectra become too sharply peaked near midrapidity. Some 
further fine-tuning of Model 2 is possible by combining the elliptic 
geometry (\ref{ellipsoid}) with an additional $\eta$-dependence of 
the freeze-out temperature ({\em Model 3}). However, the temperature 
gradients resulting from the fit are small, i.e. the width 
$\Delta\eta \gg \eta_{\max}$ is large. In spite of the considerable 
expense in computing time resulting from the introduction of the 
additional parameter $\Delta\eta$ its effect on $\chi^2$/DOF is 
only minor. 

In Figures 1--3 we compare the results of Model 3 with the 
data; the corresponding curves from Model 2 would have been hardly 
distinguishable. Remaining minor flaws of Model 3 are a slightly too 
broad rapidity spectrum and somewhat too flat $m_\T$-spectra in the 
very backward region for protons and a slight lack of curvature in the 
$m_\T$-spectra for pions. Additional pion $m_\T$-spectra covering the 
backward rapidity region $0.6 < y_{\rm lab} < 1.4$ have recently become 
available \cite{E802new}; although we did not include them in our fit
they agree with the predictions of our Model 3. Indeed, the lack of 
concavity of our calcu-
\newpage

\noindent
lated pion $m_\T$-spectra becomes less of a problem at backward rapidities.

%
%
%
 \begin{figure}
  \epsfxsize 80mm \epsfbox{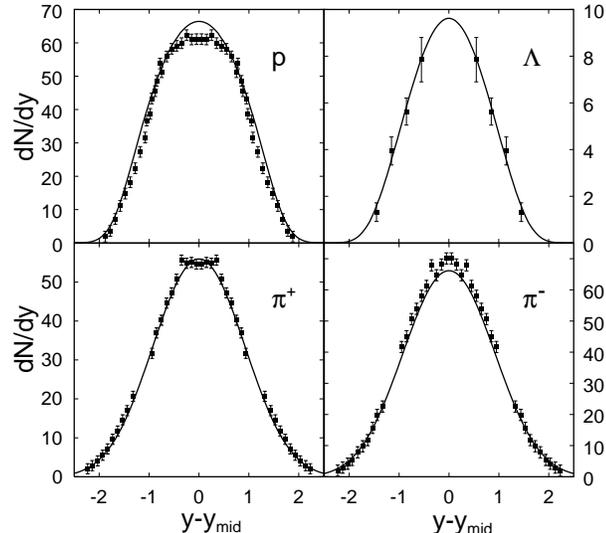}
  \caption{
      Rapidity spectra of protons, $\Lambda$ hyperons and pions in 
      Au+Au collisions at the AGS. The $\Lambda$'s were measured by 
      E891 \protect\cite{E891}. For protons and pions we use data 
      from E866 \protect\cite{E866} near midrapidity and from E877 
      \protect\cite{E877} for forward rapidities. The measured data 
      points were reflected about midrapidity using the symmetry of 
      the collision system. Solid lines: best fit with Model 3. 
 \label{F3}}
 \end{figure}
%

It is interesting to note that the thermal parameters, including
the average radial flow velocity $\langle v_\perp \rangle$,
are quite stable against the discussed variations in the model 
parametrization \cite{fn2}. Our analysis thus strongly suggests 
an average kinetic freeze-out temperature of about 90 MeV and an 
average radial flow velocity at midrapidity close to 0.5\,$c$. 
Similarly low kinetic freeze-out tempe\-ra\-tures were found in 
\cite{Nix} for the smaller Si+Au system, in a model which has 
many similarities with ours. However, the analysis of \cite{Nix} 
also included information from two-particle Bose-Einstein 
correlations which were not available for our system. The average 
transverse flow velocity at midrapidity in Si+Au is a bit smaller: 
with a linear transverse {\em velocity} profile Chapman and Nix 
\cite{Nix} found $\langle v_\perp \rangle(\eta=0) = {2\over 3} 
\times (0.683 \pm 0.048)\,c \approx 0.45\, c$.

In \cite{Stachel96} a successful fit of the rapidity spectra for pions,
kaons, protons and $\Lambda$'s with a freeze-out temperature of 130 MeV 
was reported. The rapidity spectra are, however, not very sensitive to 
the freeze-out temperature \cite{Schnedermann}; we found that the 
large body of $m_\T$-spectra are badly fit with $T=130$ MeV and clearly 
point to a considerably lower kinetic freeze-out temperature. On the 
other hand, the thermal freeze-out parameters found here correspond to
a hadron resonance gas with an entropy per baryon $S/A \simeq 12{-}13$. 
This is the same range as found in the chemical freeze-out analysis
of Si+Au collisions at the AGS \cite{Rafelski} and confirmed later
in Refs.~\cite{Braun,Stachel96,Cleymans} also for Au+Au collisions. 
Our findings are therefore consistent with chemical freeze-out at 
$T_{\rm chem} \approx 130$ MeV, followed by isentropic (hydrodynamic) 
expansion and kinetic freeze-out at $T_{\rm kin} \approx 90$ MeV. 
 
Naively one would expect that increasing beam energy leads to higher
energy deposition in the reaction zone and thus to higher temperature 
and/or stronger radial flow at freeze-out. A tendency for stronger 
flow at larger $\sqrt{s}$ is indeed seen at energies below and 
including the AGS \cite{Herrmann,Seto98}. At the higher SPS energy an 
analysis of negative particle spectra and two-particle correlations from 
the NA49 Pb+Pb experiment points to kinetic freeze-out at $T{=}100{-}120$ 
MeV with $\langle v_\perp \rangle{=}0.45{-}0.5\, c$ \cite{NA49,Tomasik}, 
the lower freeze-out temperature and larger flow being preferred 
\cite{Tomasik}. This supports our expectation, but contradicts an 
earlier analysis \cite{PBM,Herrmann} which did not differentiate between 
chemical and kinetic freeze-out and thus gave larger freeze-out 
temperatures and smaller flow velocities.

We thank J.P.~Wessels and C.A.~Ogilvie for kindly sending us their 
data files, and J. Rafelski for a very constructive remark. We gratefully 
acknowledge the support of this work by BMBF, DFG and GSI.



\begin{thebibliography}{99}
\bibitem{fn1}
  For reasons of stability of the fit the normalization constants 
  ${\cal N}_i$ were determined after fitting the thermal parameters, by
  minimizing the $\chi^2$ of the corresponding spectrum $i$ keeping 
  the other parameters fixed. This procedure is justified as long as 
  the ${\cal N}_i$ are sufficiently close to 1 such that resonance 
  feeddown is not seriously affected. Without separately optimized 
  normalization factors the best fit with Model 1 gives 
  $\chi^2$/DOF = 8.04.
\bibitem{fn2}
  For ellipsoidal freeze-out geometry the temperature and the average 
  radial flow velocity at midrapidity are somewhat higher than for
  a cylindrical fireball; once averaged over rapidity, the values are, 
  however, quite similar.
\end{thebibliography}
\end{document}